\documentstyle[12pt]{article}
\input epsf.tex

\newcommand{\be}{\begin{equation}}
\newcommand{\ee}{\end{equation}}
\newcommand{\bea}{\begin{eqnarray}}
\newcommand{\eea}{\end{eqnarray}}
\newcommand{\req}[1]{~(\ref{#1})}

\newcommand{\gs}{\mbox{$g_s$}}            
\newcommand{\ls}{\mbox{$l_s$}}            
\newcommand{\g}[1]{g_{YM_{#1}}}

\def\href#1#2{#2}

\begin{document}

\baselineskip=15.5pt
\pagestyle{plain}
\setcounter{page}{1}

\begin{titlepage}

\begin{flushright}
PUPT-1840\\
hep-th/9902197
\end{flushright}
\vspace{3mm}

\begin{center}
{\huge Baryons and Flux Tubes in Confining Gauge Theories from Brane 
Actions}
\end{center}
\vspace{5mm}

\begin{center}
{\large Curtis G.\ Callan, 
Jr.\footnote{callan@feynman.princeton.edu},
Alberto G\"uijosa\footnote{aguijosa@princeton.edu}, \\
Konstantin G. Savvidy\footnote{ksavvidi@princeton.edu} and
{\O}yvind Tafjord\footnote{otafjord@princeton.edu} }\\
\vspace{3mm}
Joseph Henry Laboratories\\
Princeton University\\
Princeton, New Jersey 08544\\
\end{center}

\begin{center}
{\large Abstract}
\end{center}
\noindent

We study baryon configurations in large $N$ non-supersymmetric
$SU(N)$ gauge theories, applying the $AdS$/CFT correspondence.
Using the D5-brane worldvolume theory in the near-horizon geometry
of non-extremal D3-branes, we find embeddings which describe
baryonic states in three-dimensional QCD. In particular, we
construct solutions corresponding to a baryon made of $N$ quarks,
and study what happens when some fraction $\nu$ of the total
number of quarks are bodily moved to a large spatial separation
from the others. The individual clumps of quarks are represented
by Born-Infeld string tubes obtained from a D5-brane whose
spatial section has topology ${\bf R}\times {\bf S}^4$.
They are connected by a confining color flux tube, described
by a portion of the fivebrane that runs very close and parallel
to the horizon. We find that this flux tube has a tension with a
nontrivial $\nu$-dependence (not previously obtained by other 
methods).
A similar picture is presented for the four-dimensional case.

\vfill
\begin{flushleft}
February 1999
\end{flushleft}
\end{titlepage}
\newpage

\section{Introduction}

The Born-Infeld action plus an appropriate Wess-Zumino term defines
a worldvolume theory for D-branes which has proved to be a
powerful way to describe these objects and their excitations.
In the context of Maldacena's correspondence between supergravity
in anti-de Sitter ($AdS$) space and certain gauge theories
\cite{jthroat,gkpads,wittenholo},
there are gauge theory questions which can be answered by the study
of branes extended in curved space. In particular, it was shown in a
general way that the large-$N_c$ dynamics of baryons can be related 
to the
behavior of D5-branes extended in $AdS$ space \cite{wittenbaryon, 
groguri}.
Concrete realizations of this possiblity in the context of the 
non-confining
${\mathcal N}=4$ supersymmetric gauge theory have been worked out
in \cite{cgs,imamura}
using the Born-Infeld approach for constructing strings out of 
D-branes
\cite{calmal,gibbons}. In this paper, we will look at how these 
constructions
work in the more complicated spacetimes that correspond to various 
confining
gauge theories. We will examine confining forces by looking at what 
happens
to baryons when they are pulled apart into their quark constituents. 
This
will be compared to (and yields somewhat more information than) the 
study
of confining forces via simple strings that `hang' into the $AdS$ 
geometry
from a boundary Wilson loop \cite{adswilson, bisy2, groguri}.

In this paper, we extend our previous work \cite{cgs} on D5-branes in
extremal background in three ways. First, we will study non-extremal
supergravity backgrounds, corresponding to gauge theories 
dimensionally
reduced in a way that breaks supersymmetry. Second, we allow the 
brane
configurations to have extension in the spacetime coordinates of the
gauge-theory instead of being localized at a point. This will allow 
us
to describe a baryon which is being `pulled apart' into quark 
constituents.
Third, we discuss also the case based on D4-branes, which corresponds 
to
a gauge theory in four dimensions.

We start out in Section \ref{3dsec} by analysing baryons in a 
(2+1)-dimensional $SU(N)$ Yang-Mills theory which is obtained from 
(3+1)-dimensions by compactifying on a supersymmetry breaking circle.
As proposed in \cite{wittenthermo}, this gauge theory is dual to
a certain non-extremal D3-brane geometry\footnote{See \cite{klebanov,
minahan, others} 
for 
some interesting alternative string theory approaches 
to the study of large-$N$ non-supersymmetric Yang-Mills theories.} 
and, following \cite{cgs},
we study solutions of the D5-brane worldvolume equations of motion
in that geometry. We find a class of solutions that are localized in
the gauge theory spatial coordinates and appear to describe the
baryon. Unlike the baryons constructed in the extremal background
\cite{cgs,imamura}, these solutions have no moduli since the quarks
are truly bound in the baryon. We then study a new class of solutions
in which the $N$ quarks are separated into two groups, containing
$\nu N$ and $(1 - \nu)N$ quarks respectively, separated by a spatial
distance $L$ in the gauge theory. The $L$-dependence of the energy of
these solutions is consistent with confinement and the implied color
flux tube tension has a non-trivial dependence on the color charge 
$\nu$.

In Section \ref{4dsec} we carry out the same analysis for the
gauge theory in one more dimension, i.e., D4-branes embedded in a
non-extremal D4-brane geometry. We include also a short section on
the extremal limit, as this was not covered in \cite{cgs}. The
analysis is parallel to the one in Section \ref{3dsec}, with the
added surprise that the resulting color flux tension now has a
very simple dependence on $\nu$.

In a final section we summarize our results and discuss 
future directions.

\section{The Baryon in Three Dimensions} \label{3dsec}

\subsection{Worldvolume Action and Equations of Motion}
\label{eqn3dsec}

We derive the equations for a D5-brane embedded in the near-horizon
geometry of $N$ nonextremal D3-branes. The Euclidean background 
metric is
\bea \label{d3metric}
{}&{}&ds^2=\left(\frac{r}{R}\right)^{2}\left[f(r)d\tau^2+dx_{||}^2
\right]
+\left(\frac{R}{r}\right)^{2}f(r)^{-1}dr^2+R^2 d\Omega_5^2, \\
{}&{}&f(r)=1-r_{h}^4/r^4, \qquad R^{4}=4\pi\gs N \ls^{4}, \qquad
r_{h}=\pi R^{2} T, \nonumber
\eea
where $\{\tau,x_{||}\}=\{\tau,x,y,z\}$ denote the directions parallel 
to the
threebranes. The coordinate $\tau$ is periodic, with period
$1/T$, where $T$ is the Hawking temperature.
The relation between the horizon radius $r_{h}$ and $T$ ensures 
smoothness
of the geometry at $r=r_{h}$.

Under the $AdS$/CFT correspondence \cite{jthroat,gkpads,wittenholo},
type IIB string theory on a background with the above metric, a
constant dilaton, and $N$ units of fiveform flux through the 
five-sphere,
is dual to ${\mathcal N}=4$, $d=3+1$ $SU(N)$ SYM theory at 
temperature
$T$, with coupling $\g{4}^{2}=2\pi\gs$. The gauge theory coordinates
are $\{x,y,z,\tau\}$. For large $T$ the $\tau$
circle becomes small and one effectively obtains a description of a
strongly-coupled (coupling $\g{3}^{2}=\g{4}^{2}T$) three-dimensional
Euclidean gauge theory at zero temperature. The thermal boundary
conditions on the circle break supersymmetry and the fermions and
scalars acquire masses of order $T$ and $\g{4}^{2}T$, respectively.
The effective three-dimensional theory is expected to display 
behavior
similar to that of non-supersymmetric pure Yang-Mills theory,
$\mbox{QCD}_{3}$ \cite{wittenthermo}.

A baryon (a bound state of $N$ external quarks) in the 
three-dimensional
theory has as its string theory counterpart a fivebrane wrapped on an
${\bf S}^{5}$ on which $N$ fundamental strings terminate
\cite{wittenbaryon,groguri}. The fivebrane worldvolume action is
$$
S = -T_5 \int d^6\xi\sqrt{\det(g+F)} +T_5 \int A_{(1)}\wedge 
G_{(5)}~,
$$
where $T_5=1/(\gs(2\pi)^{5}\ls^{6})$ is the brane tension.
The Born-Infeld term involves
the induced metric $g$ and the $U(1)$ worldvolume
field strength $F_{(2)}=d A_{(1)}$.
The second term is the Wess-Zumino coupling of the
worldvolume gauge field $A_{(1)}$ to (the pullback of)
the background five-form field strength $G_{(5)}$, which effectively
endows the fivebrane with a $U(1)$ charge proportional to the
${\bf S}^{5}$ solid angle that it spans.

For a static baryon we need a configuration invariant under 
translations
in the gauge theory time direction, which we take to be $y$. We use 
$y$
and the ${\rm \bf S}^{5}$ spherical coordinates as worldvolume 
coordinates
for the fivebrane, $\xi_{\alpha}=(y,\theta_{\alpha})$. For simplicity 
we
restrict our attention to $SO(5)$ symmetric configurations of the 
form
$r(\theta)$, $x(\theta)$, and $A_y(\theta)$ (with all other fields 
set to
zero), where $\theta$ is the polar angle in spherical coordinates.
The action then simplifies to
\be \label{d3action}
S= T_5 \Omega_{4}R^4\int dy\,d\theta \sin^4\theta \{ -
  \sqrt{r^2+r^{\prime 2}/f(r)+(r/R)^{4}x^{\prime 2}-F_{y\theta}^2}
  +4 A_y \},
\ee
where $\Omega_{4}=8\pi^{2}/3$ is the volume of the unit four-sphere.

The gauge field equation of motion following from this action reads
$$
\partial_\theta D(\theta) = -4 \sin^4\theta,
$$
where the dimensionless displacement $D$ is the variation of the
action with respect to $E=F_{y\theta}$. The solution to this equation 
is
\be \label{d}
D(\theta) = \left[{3\over 2}(\nu\pi-\theta)
  +{3\over 2}\sin\theta\cos\theta+\sin^{3}\theta\cos\theta\right].
\ee
As will be explained below, the integration constant $0\leq\nu\leq 1$
controls the number of Born-Infeld strings emerging from each pole of
the ${\bf S}^{5}$. Next, it is convenient to eliminate the gauge 
field
in favor of $D$ and Legendre transform the original Lagrangian to
obtain an energy
functional of the embedding coordinate $r(\theta)$ only:
\be \label{u}
U =  T_5 \Omega_{4}R^4\int d\theta
\sqrt{r^2+r^{\prime 2}/f(r) +(r/R)^{4}x^{\prime 2}}\,
\sqrt{D(\theta)^2+\sin^8\theta}~.
\ee
This action has the interesting scaling property that if
$\{r(\theta),x(\theta)\}$ is a solution for horizon
radius $r_{h}$, then $\{\alpha r(\theta), \alpha^{-1}x(\theta)\}$
is a solution for horizon radius $\alpha r_{h}$. The scaling
$x\propto R^{2}/r$ is precisely as expected from the
holographic UV/IR
relation \cite{susswi,pp}. We will have more
to say about scaling behavior of solutions later on.

The fivebrane embeddings of interest to us will have singularities:
places on the five-sphere (typically $\theta\to \pi$ or $0$) where
$r\to\infty$ and $x^{\prime}\to 0$. As in~\cite{cgs,calmal,gibbons},
these `spikes' must be interpreted as bundles of fundamental strings
attached to the wrapped fivebrane and localized at some definite 
value of $x$.
It can be seen from\req{u} that a spike sticking out at $\theta=\pi$ 
has a
`tension' (energy per unit radial coordinate distance)
$T_5 \Omega_{4} R^4 |D(\pi)|f(r)^{-1/2}=(1-\nu)N T_F f(r)^{-1/2}$,
which is precisely the tension of $(1-\nu)N$ fundamental strings in
this geometry. A spike at $\theta=0$ has the same tension as $\nu N$
strings so that, taken together, the two singularities
represent a total of $N$ fundamental strings, as expected. Surfaces
with more singularities and less symmetry are perfectly possible, but
a lot harder to analyze. To keep things manageable, we have built
$SO(5)$ symmetry into the ansatz.

In the extremal case ($r_{h}=0$) discussed in~\cite{cgs}, the
BPS condition provided a first integral which greatly simplified the
analysis. In the nonextremal case we are now discussing, there is
no such first integral and we have to deal with the unpleasant
second order Euler-Lagrange equation that follows from\req{u}. This
is most conveniently done in a parametric Hamiltonian formalism
\footnote{We would like to thank G.~Savvidy for suggesting and 
helping
to realize this approach.}. First we rewrite the energy in terms of a 
general
worldvolume parameter $s$ with the D5-brane embedding defined by
functions $\theta=\theta(s)$, $r=r(s)$, $x=x(s)$:
\be \label{upar}
U = T_5 \Omega_{4}R^4\int ds
\sqrt{ r^2\dot{\theta}^2 + \dot{r}^2/f+(r/R)^{4}\dot{x}^2}~
\sqrt{D^2+\sin^8\theta},
\ee
where dots denote derivatives with respect to $s$.
The momenta conjugate to $r$, $\theta$ and $x$ are
\be \label{mom}
p_r=f^{-1}\dot{r}\Delta, \quad
p_{\theta}=r^2\dot{\theta}\Delta, \quad
p_{x}=(r/R)^{4}\dot{x}\Delta, \quad
\Delta=\frac{\sqrt{D^2+\sin^8\theta}}
   {\sqrt{ r^2\dot{\theta}^2 + \dot{r}^2/f+(r/R)^{4}\dot{x}^2}}~.
\ee
The Hamiltonian that follows from the action\req{upar} vanishes 
identically
due to reparametrization invariance in $s$. Furthermore, the momentum
expressions are non-invertible and the system is subject to the 
constraint
\be \label{ham}
2\tilde{H} =
    \left(1-\frac{r_h^4}{r^4}\right) p_r^2 + \frac{p_{\theta}^2 
}{r^2}
+\frac{R^4}{r^4}p_{x}^2-
\left( D^2+\sin^8\theta \right) =0~.
\ee
This constraint can be taken as the Hamiltonian and this choice
conveniently fixes the gauge, while getting rid of the
complicated square-root structure of the action.
The equations of motion that follow from this Hamiltonian are

\hfill
\parbox{1.7in}
{\begin{eqnarray*}
\dot r &=&\left(1-\frac{r_h^4}{r^4}\right) p_r~,\\
\dot\theta &=&\frac{p_\theta}{r^2}~, \\
\dot x &=& \frac{R^{4}}{r^{4}}p_{x},
\end{eqnarray*}}
\parbox{3.1in}
{\begin{eqnarray*}
\dot p_r &=&\frac{2}{r^5}(p_x^2 R^4 -p_r^2 r_h^4)
     + \frac{p_\theta^2}{r^3}~, \\
\dot p_\theta &=& -6\sin^4\theta \left(\pi\nu-\theta+ 
\sin\theta\cos\theta
\right)~,\\
\dot{p}_{x}&=& 0~.
\end{eqnarray*}}
\hfill
\parbox{0.5in}
{\bea \label{eom}
\eea}
Together with initial conditions, these equations completely define
the solutions for the fivebrane. The initial conditions should be
chosen such that $\tilde H=0$.

\begin{figure}[htb]
\centerline{\epsfxsize=14cm
\epsfbox{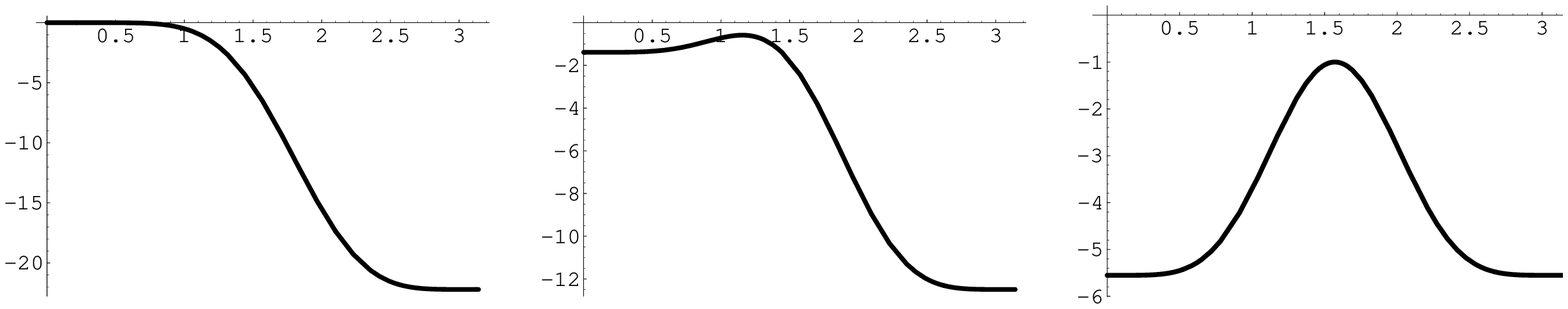}}
\caption{\small The potential
  $V(\theta)=-\left[D(\theta)^2+\sin^8\theta\right]$ for
  $\nu=0,1/4,1/2$ (see text for discussion).}
\label{D3pot}
\begin{picture}(0,0)
\put(12,41){\scriptsize $V(\theta)$}
\put(53,37){\scriptsize $\theta$}
\put(19,20){\small $\nu=0$}
\put(58,41){\scriptsize $V(\theta)$}
\put(98.5,37){\scriptsize $\theta$}
\put(65,20){\small $\nu=1/4$}
\put(102,41){\scriptsize $V(\theta)$}
\put(144,38){\scriptsize $\theta$}
\put(117,20){\small $\nu=1/2$}\end{picture}
\end{figure}

To gain some insight into the solutions to these equations, notice 
that
the basic problem to solve is a motion in the two-dimensional 
$r-\theta$
plane: the motion in $x$ is then determined by the choice of a 
conserved
value for $p_x$. Note that for $p_x=0$, the surface sits at a fixed 
value
of $x$ and therefore has no spatial extension in the gauge theory
coordinates: we will call this a `point' solution. When $r$ is large 
compared
to $r_{h}$ and $R \sqrt{p_{x}}$, 
the $r-\theta$ motion is simply that of a particle
of unit mass moving in two dimensions under the influence of the 
angular
potential $V(\theta)=-\left[D(\theta)^2+\sin^8\theta\right]$.
By the constraint, the energy of this fictitious particle vanishes.
For generic $\nu$, the potential has three extrema (see Figure 
\ref{D3pot}):
two minima at $\theta=0$ and $\theta=\pi$, and a maximum at
$\theta=\theta_{c}$ such that 
$\pi\nu=\theta_{c}-\sin\theta_{c}\cos\theta_{c}$.
For large $r$ the particle will thus roll down towards one of the two 
minima.
Whether it reaches $\theta=0,\pi$ at a finite value of $r$ depends on
the initial boundary conditions.

\subsection{The Point Baryon}\label{point3dsec}

In this section we study solutions which correspond to a baryon
localized at a particular gauge theory position. To localize the 
surface
in $x$, we just set $p_{x}=0$.
With the symmetry that we have built in, the equations of motion 
typically
allow the surface to run off to $r=\infty$ at $\theta=\pi$ or $0$.
At least asymptotically, such `spikes' are equivalent to bundles of
fundamental strings and will be identified with the quark 
constituents
of the state represented by the wrapped fivebrane. To get a baryon 
whose component quarks have identical $SU(4)$ (flavor) quantum 
numbers,
we want a spike representing $N$ quarks to emerge from one pole of 
the
$S^5$ (say $\theta=\pi$) with a smooth surface at the other pole.
To meet the first condition, it suffices to set the integration 
constant
$\nu=0$ and to meet the second, we impose smooth boundary conditions
($\partial_{\theta}r=0$ and $r=r_{0}$) at $\theta=0$.\footnote{This
is equivalent to requiring $p_{r}=0$ at
$\theta=0$, in which case $p_{\theta}$ must also vanish to satisfy 
the
constraint\req{ham}.}

\begin{figure}[htb]
\centerline{\epsfxsize=12cm
\epsfbox{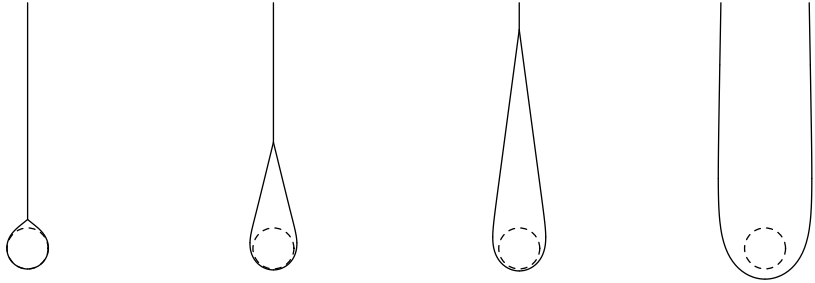}}
\caption{\small Family of solutions illustrating the progressive
 deformation of the fivebrane by the bundle of $N$ fundamental
 strings. The dotted  circle represents the horizon.
 The stable configuration is a `tube' with
 $r_{0}\to\infty$.}
\label{D3tubes}
\end{figure}

Numerical integration with these boundary conditions yields a
one-parameter family of solutions (parametrized by $r_0$). Due to the
scaling mentioned before, the solutions really only depend on
$r_{h}$ through the ratio $\zeta=r_{0}/r_{h}\ge 1$.
Figure \ref{D3tubes} shows polar plots\footnote{Although these plots
provide a conveniently simple representation of the solutions,
the reader should bear in mind that they are a bit misleading as to 
the
intrinsic geometry, since the radius of the ${\bf S}^{5}$ is $R$,
independent of $r$.} of the solutions for a few representative
values of $\zeta$. When $\zeta<\zeta_{crit}\approx 9$ the solution 
tilts
toward and eventually crosses the symmetry axis,
thus reaching $\theta=\pi$ at a finite value of $r$. As $\zeta\to 1$
(i.e. as the starting radius gets closer
and closer to the horizon),
the brane becomes more and more spherical (notice that
$r(\theta)=r_{h}$ is a solution to the equations of motion).
For $\zeta>\zeta_{crit}$, on the other hand,
the solution tilts away from the symmetry
axis, reaching $\theta=\pi$ only at $r=\infty$.
As $\zeta$ increases the solution looks more and more like a
`tube'. This is simply a consequence of the fact that, for large
$\zeta$, the solution is always very far from the horizon, and is
well-approximated by an extremal embedding (discussed in \cite{cgs}).
When $\zeta\to\infty$ the configuration
becomes a BPS `tube' \cite{cgs} with infinite radius.

The cusp in the $\zeta<\zeta_{crit}$ solutions indicates the presence 
of a
delta-function source in the equations of motion. Since the compactly
wrapped brane intercepts $N$ units of five-form flux, it has $N$ 
units
of worldbrane $U(1)$ charge and must have $N$ fundamental strings
attached to it \cite{wittenbaryon}. This is
most simply achieved by taking the cusp as the point of attachment of
$N$ fundamental strings, running along the ray $\theta=\pi$.
In accordance with the Born-Infeld string philosophy 
\cite{calmal,gibbons},
these strings are equivalent to a D5-brane wrapped on an $S^{4}$ of 
vanishing
volume which carries the $U(1)$ flux out to infinity. A simple 
modification
of the flat space argument \cite{emp,calmal,gibbons} shows that such 
a
collapsed fivebrane is a solution to the equations of motion and has
`tension' $N T_Ff(r)^{-1/2}$ (exactly the tension exerted by a
fundamental string of intrinsic tension $T_F$ in this curved space).

The entire fivebrane-string (or collapsed brane) system will be
stable only if there is tension balance between its two components.
To obtain the stability condition, let $r_{c}$ denote the location of
the cusp (which is a function of $r_{0}$), and
parametrize the family of fivebrane
embeddings as $r=r(\theta; r_{c})$. Under the variation $r_{c}\to
r_{c}+\delta r_{c}$, it can be seen from\req{u}, after an
integration by parts and application of the Euler-Lagrange equation,
that the energy of the brane changes only by a surface term,
\be \label{varcalc}
\frac{\partial U}{\partial r_{c}}=T_5 \Omega_{4}R^4
    \frac{r^{\prime}\sqrt{D^2+\sin^8\theta}}
    {f\sqrt{r^2+f(r)^{-1}r^{\prime 2}}}
    \left.\frac{\partial r}{\partial r_{c}}\right|^{\pi}_0
 =\frac{NT_Ff(r_{c})^{-1/2}}{\sqrt{1+f(r_{c}) r_{c}^2/r_{c}^{\prime
 2}}}~,
\ee
where $r_{c}^{\prime}=\partial_{\theta}r|_{\theta=\pi}$, and
we have used the fact that $r(\pi;r_{c})=r_{c}$. The numerator
in the last expression of\req{varcalc} is the `tension' at $r=r_{c}$ 
of $N$
fundamental strings, so it is clear that the brane has a
lower tension for any $r_{c}>r_{h}$. The energy is lowered by 
expanding
the fivebrane and shortening the explicit fundamental string.
A similar variational calculation applied to the $\zeta>\zeta_{crit}$
configurations (cut off at a large $r=r_{max}$) shows
that the BPS `tube' at infinity is the lowest energy solution.
This is consistent with the results of \cite{baryonsugra,imafirst},
where the baryon was examined using the pure Nambu-Goto action for 
the
fivebrane wrapped on a sphere. We emphasize that the above 
variational
calculation used solutions of the full Born-Infeld (plus
Wess-Zumino) action.

Altogether, then, the solutions depicted in Fig.~\ref{D3tubes} 
provide
a physically satisfying picture of the process through which the $N$
fundamental strings deform the initially spherical fivebrane, pulling
it out to infinity. The final configuration has the shape of a 
`tube',
just like the BPS embeddings found in~\cite{imamura,cgs}. In the
supersymmetric case, $r_{0}$ was a modulus and the energy of the
baryon was independent of the overall scale of the solution.
In the nonextremal case examined here, however, there
appears to be a potential for that modulus which drives the stable 
solution
out to $r_{0}\to\infty$.

The dependence of the fivebrane embedding on the $S^{5}$ coordinates
encodes the flavor structure (i.e., the $SU(4)$ quantum numbers) of
the gauge theory state under consideration. As a result of the UV/IR
relation, the $AdS$ radial coordinate $r$ is associated with an 
energy scale
in the gauge theory, $E=r/R^{2}$ \cite{susswi,pp}. The embedding
$r(\theta)$ consequently associates a particular value of $\theta$ to
each different distance scale, yielding some sort of $SU(4)$ 
wavefunction for
the baryon. The $SO(5)$ symmetry of the embedding translates into the
statement that the baryon is a singlet under the corresponding
$SU(4)$ subgroup. Finally, the fact that a given surface spans
the range $r\ge r_{0}$ implies that the dual gauge theory 
configuration
has structure on all length scales from zero up to a characteristic 
size
$R^{2}/r_{0}$. Since the energetically preferred configuration has 
$r_{0}\to\infty$, it is in this sense truly pointlike.

\subsection{The Split Baryon: Color Dependence of the String Tension}
\label{split3dsec}

We now turn our attention to solutions with $p_{x}\neq 0$
(i.e. $x^{\prime}\neq 0$). They describe collections of quarks
at finite separation in the gauge theory position space and are of
interest for exploring confinement issues. It turns out to
be rather easy to construct a surface describing an $SU(N)$
baryon split into two distinct groups, containing $\nu N$ and
$(1-\nu)N$ quarks respectively and separated by a distance $L$
in the $x$ direction. In a
confining $SU(N)$ gauge theory, two such quark bundles should be
connected by a color flux tube and we will study the Born-Infeld
representation of this phenomenon. Each group of quarks corresponds
as before to a bundle of Born-Infeld strings, realized in our
approach as a singular spike or fivebrane `tube' with topology
${\bf R}\times {\bf S}^{4}$. Remember that we have assumed an
$SO(5)$-symmetric configuration, which means that the two
singularities representing the quarks must be located at opposite
poles of the ${\bf S}^{5}$ (we will put them at $\theta=0$ and
$\theta=\pi$) with corresponding implications about the $SU(4)$
flavor structure of the states we are constructing. More general
flavor structures are possible, but we will not try to study these
more complicated surfaces. For large spatial separation
$L=\vert x(\infty)-x(-\infty)\vert$, the portion of the fivebrane 
that
interpolates between the two string bundles runs close to the
horizon and it is this part of the surface that encodes the
confining flux tube of the gauge theory. The surface equations 
\req{eom}
imply that the part of the surface that has large spatial extent
must sit at a constant $\theta=\theta_{c}$ where $\dot p_{\theta}=0$.
More precisely, it has to sit at the solution of
\be \label{nu}
\pi\nu=\theta_{c}-\sin\theta_{c}\cos\theta_{c}~.
\ee
corresponding to the unstable maximum of the
potential $V(\theta)$ discussed at the end of Section \ref{eqn3dsec}.
The critical angle is a monotonic function of $\nu$, such that
$\theta_{c}(0) =0$ and $\theta_{c}(\nu)=\pi-\theta_{c}(1-\nu)$.
The energetics of the part of the fivebrane that encodes the 
confining
flux tube will depend on $\theta_c$, and therefore $\nu$, in a way 
that
we will now examine in some detail.

\begin{figure}[htb]
\centerline{\epsfxsize=6cm
\epsfbox{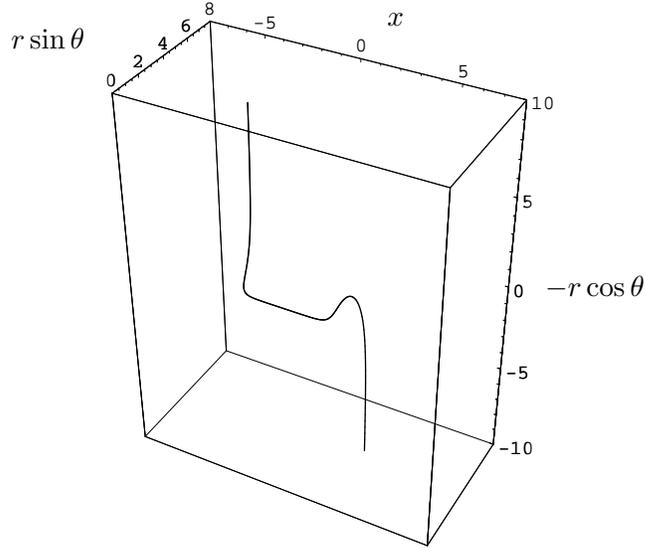}}
\begin{picture}(0,0)
\put(35,72){\small $r\sin\theta$}
\put(85,75){\small $x$}
\put(106,39){\small $-r\cos\theta$}
\end{picture}
\caption{\small The three-dimensional projection of the D5-brane.
Every point on the curve is an ${\bf S}^{4}$.
One can clearly see how the brane drops down towards the horizon,
extends horizontally along it, and finally leaves at the other end.
{}From the point of view of the three-dimensional $SU(N)$ gauge 
theory,
which lives in the $\{x,y,z\}$ directions, this configuration
represents a baryon split into two groups of
$\nu N$ and $(1-\nu)N$ quarks (the vertical segments
--- see Fig.~\ref{splitpol}),
connected by a flux tube extending a finite distance along the $x$
direction.
}
\label{split3d}
\end{figure}

\begin{figure}[htb]
\centerline{\epsfxsize=6cm
\epsfbox{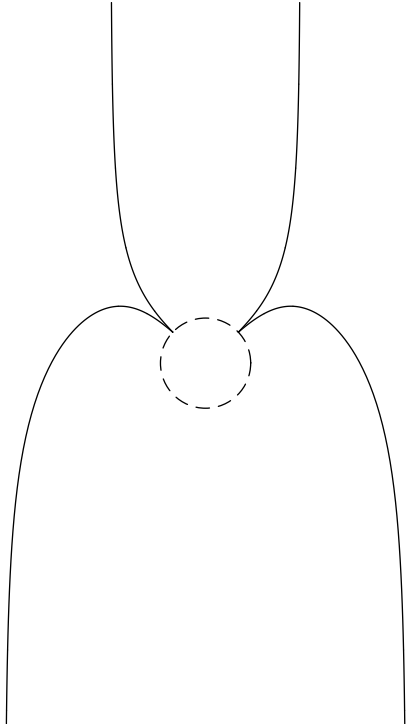}}
\caption{\small Polar plot of $r(\theta)$ -- the two-dimensional
projection of Fig.~\ref{split3d}.
The Born-Infeld string
`tubes' pointing up and down represent two groups of $\nu N$ and 
$(1-\nu)N$
quarks, respectively (with $\nu=0.9$ here).
The brane extends in the $x$-direction mostly
at the inflection point (the cusp), while
$x$ is essentially constant along the tubes.}
\label{splitpol}
\end{figure}

Unfortunately, we must resort to numerical analysis to construct 
specific
surfaces of this kind. It is convenient to take the point of closest 
approach to the horizon as the starting point for the numerical 
integration. 
We start the integration off with the initial conditions
\be \label{inicon}
\begin{array}{ll}
r(0)= r_h + \epsilon~,&  p_r(0)=0~,\\
\theta(0)=\theta_{c}~,&  p_{\theta}(0)=\eta~,\\
x(0)=0~,&  p_x(0)=(r(0)/R)^{2}\sqrt{\sin^6\theta_{c}
-(\eta/ r(0))^2}~.
\end{array}
\ee
The distance from the horizon at the point of closest approach is 
controlled
by $\epsilon$. For a given $\epsilon$, we have to `shoot' in $\eta$ 
until
we get
satisfactory behavior of the quark-like singularities at
$\theta\rightarrow 0$ and $\theta \rightarrow \pi$ (see Section
\ref{point3dsec} for details). Indeed, it is natural to require
asymptotically BPS behavior in the region of space
where supersymmetry is recovered locally (i.e., far from the 
horizon).
Once that is done, $\epsilon$ controls the spatial separation of the 
two
separated quark bundles. Figs.~\ref{split3d} and \ref{splitpol}
depict a typical fivebrane embedding obtained by numerical 
integration,
for the case $\nu=0.9$~. It can be seen in Fig.~\ref{split3d} that 
the brane
extends in the $x$ direction mostly in its `flux tube' portion, at
$\theta=\theta_{c}$ and $r\approx r_{h}$. The Born-Infeld string 
`tubes'
corresponding to the two groups of quarks lie essentially at a 
constant
value of $x$. 

It is seen from the equation for $\dot r$ in (\ref{eom}) that, for 
the
portion of the brane running close to the horizon, $r(s)-r_h$
grows as an exponential in the parameter $s$,
with an exponent proportional to $p_r$. The latter
reaches a value close to $(R/r_{h})p_x$ (see the equation for $\dot 
p_r$).
The equation for $\dot x$ then shows that the separation $L$
between the quarks increases only logarithmically with $\epsilon$, 
the
minimal distance to the horizon. In fact, there
exists a limiting solution which consists of $\nu N$ quarks with a 
flux tube
that extends to infinity, and the brane approaches the horizon 
exponentially
with distance.

{}From the above discussion it is clear that for large quark 
separation $L$
the (renormalized) energy will receive its main contribution from the 
flux
tube, and will consequently depend linearly on $L$, a clear 
indication of 
confinement.  It is easy to compute the tension (energy per unit 
distance in $x$) of the color flux tube.
Note that when the fivebrane runs parallel
and close to the horizon, the energy function (\ref{u}) reduces
to
$$U_{flux}=\frac{2}{3\pi} N T_{F}
   \int dx \left(\frac{r}{R}\right)^{2}
    \sqrt{D^2 + \sin^8 \theta}~. $$
Using $r\simeq r_{h}$, $\theta\simeq\theta_{c}$, and performing some 
simple
manipulations using the definitions of $\theta_c$\req{nu} and 
$D(\theta)$\req{d}, one obtains the tension
\be \label{tension}
\sigma_{3}(\nu)=\frac{2}{3\pi}N T_{F} \left(\frac{r_h}{R}\right)^{2}
  \sin^{3}\theta_{c}=\frac{\sqrt{2}}{3}N\sqrt{\g{3}^{2}N}T^{3/2}
  \sin^{3}\theta_{c}~,
\ee
where the last expression is given solely in terms of parameters of
the gauge theory in three dimensions. Since it is obtained by
dimensional reduction from four dimensions, this theory should be
understood to have an ultraviolet cutoff proportional to
the Hawking temperature $T$.
The dependence of the tension on $T$, $N$, and the 't~Hooft
coupling $\lambda_{3}=\g{3}^{2}N$ agrees with the result
of~\cite{baryonsugra}, where a baryon whose component quarks lie on
a circle is treated within a simplified
Nambu-Goto approach.

Notice that, in addition, equation\req{tension}
gives the dependence of the flux tube tension
on $\nu$, i.e. on its color content. This nontrivial 
dependence, arising entirely from the factor
$\sin^{3}\theta_{c}$, is plotted in Fig.~\ref{D3tension}:

\vspace{0.4cm}

\begin{figure}[htb]
\centerline{\epsfxsize=8cm
\epsfbox{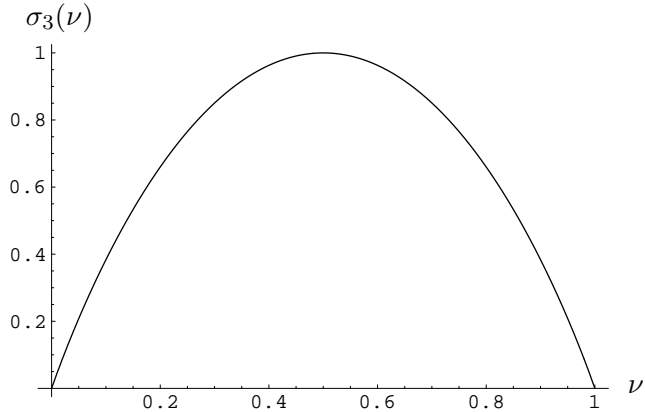}}
\begin{picture}(0,0)
\put(40,57){\small $\sigma_3(\nu)$}
\put(120,8){\small $\nu$}
\end{picture}
\caption{\small The tension of the flux tube (normalized to unity
 at its peak) as a function of
 $\nu$, the fraction of quarks pulled apart.
 In the full theory $\nu$ should be quantized in units of ${1/N}$.
  See text for discussion.}
\label{D3tension}
\end{figure}

Let us discuss the features of the flux tube
tension seen in Fig.~\ref{D3tension}.
As one might expect, the tension increases linearly for small $\nu$.
This means that for each additional quark removed the work done is
approximately constant.
Also as expected, in expression\req{tension} there is a
complete symmetry between $\nu$ and $1-\nu$, (i.e., it makes no 
difference
whether $n$ or $N-n$ quarks are pulled out). Thus, the
tension has a maximum
at $\nu=1/2$ and comes back down to zero near $\nu=1$. In gauge 
theory
language,
the flat part of the curve means that it does not cost any energy to 
move
the quark from one bundle with $\sim N/2$ quarks to the other.

Notice from equation\req{nu} that $\sin^{3}\theta_{c}\simeq 
3\pi\nu/2$
for small $\nu$. This implies that $\sigma_{3}(\nu=1/N)$ becomes {\em
independent} of $N$ in the 't~Hooft limit $N\to\infty$ with
$\lambda_{3}$ fixed.  This result has a natural
gauge theory interpretation.  When one quark is pulled out from the
$SU(N)$ baryon (a color-singlet), the remaining $N-1$ must be in the
anti-fundamental representation of the gauge group.  The flux tube
extending between this bundle and the solitary quark should then have
the same properties as the standard QCD string which connects a quark
and an antiquark.  In particular, its tension should depend on $N$
only through the 't~Hooft coupling, as we have found. As a matter of
fact, for $\nu=1/N\ll 1$ equation\req{tension} precisely agrees with
the quark-antiquark string tension which follows from a Nambu-Goto
string calculation \cite{bisy2}.  More generally, for $\nu=n/N$, with
$n$ fixed as $N\to\infty$, expression\req{tension} reduces to the
tension of $n$ quark-antiquark strings.

It is important to note that, as has been pointed out by various 
authors,
the gauge theory under study here is not strictly
three-dimensional \cite{bisy2, groguri}.
The energy scale associated with the QCD string
tension, for instance, is proportional to $\lambda_{4}^{1/4}T$, where 
$\lambda_{4}=\g{4}^{2}N$.
This is much larger than the
compactification scale $T$ in the large $\lambda_{4}$ regime where
the supergravity approximation is appropriate.

\section{The Baryon in Four Dimensions} \label{4dsec}

We are of course in principle more interested in baryons in
four-dimensional gauge theories. One way of getting at this
problem is via the dynamics of a D4-brane embedded in the
background of a large number, $N$, of nonextremal D4-branes 
\cite{groguri,baryonsugra,alisha}.
The argument is completely analogous to the D5-brane in a multiple
D3-brane background that has been analysed in previous sections.

\subsection{Worldvolume Action and Equations of Motion}
\label{eqn4dsec}

The near-horizon metric and dilaton field in the D4-brane background 
are\footnote{We thank N.~Itzhaki for pointing out a numerical
error in the expression for $r_h$ that appeared in a previous version
of this paper.}
\bea \label{d4metric}
{}&{}&ds^2=\left(\frac{r}{R}\right)^{3/2}\left[f(r)d\tau^2+dx_{||}^2
\right]
+\left(\frac{R}{r}\right)^{3/2}f(r)^{-1}dr^2+R^{3/2}\sqrt{r}
d\Omega_4^2, \nonumber\\
{}&{}&e^{\phi}=\gs\left(\frac{r}{R}\right)^{3/4}, \\
{}&{}&f(r)=1-r_{h}^3/r^3, \qquad R^{3}=\pi\gs N \ls^{3}, \qquad
r_{h}=\frac{16\pi^2}{9} R^{3} T^{2}, \nonumber
\eea
where $\{\tau,x_{||}\}=\{\tau,x,y,z,w\}$ denote the directions 
parallel to the
four-branes. The Hawking temperature $T$ is also the inverse
period of the Euclidean coordinate $\tau$.
The string coupling constant is
determined by the dilaton in the original asymptotically
flat region, $\gs=e^{\phi_{\infty}}$. The IIA supergravity
solution\req{d4metric} can only be trusted in the region where both 
the
curvature in string units and the dilaton are small,
\be \label{d4valid}
1\ll  R^{3}r\ls^{-4} \ll N^{4/3}.
\ee

Type IIA string theory on this
background (which can be understood as a compactification of an 
M5-brane
system)
is dual to $d=4+1$ $SU(N)$ SYM theory (the infrared limit
of the (0,2) theory on a circle) at temperature
$T$, with coupling $\g{5}^{2}=(2\pi)^{2}\gs\ls$ \cite{juandp}.
The gauge theory coordinates
are $\{x,y,z,w,\tau\}$. For large $T$ the $\tau$
circle becomes small and the theory is effectively a strongly-coupled
four-dimensional Euclidean gauge theory at zero temperature, with 
coupling
$\g{4}^{2}=\g{5}^{2}T$,
expected to display behavior similar to that of non-supersymmetric 
pure
Yang-Mills theory, $\mbox{QCD}_{4}$ \cite{wittenthermo, groguri}.

The baryon in the
four-dimensional theory is dual to a string theory
fourbrane on which $N$ fundamental strings terminate
\cite{wittenbaryon, groguri}.
The worldvolume action for the fourbrane is
$$
S = -T_4 \int d^5\xi e^{-\tilde{\phi}}\sqrt{\det(g+F)}
  +T_4 \int A_{(1)}\wedge G_{(4)}~,
$$
where $T_4=1/(g_{s}(2\pi)^{4}\ls^{5})$ is the brane tension, and
$\tilde{\phi}=\phi-\phi_{\infty}$.
The Wess-Zumino term couples the
worldvolume gauge field $A_{(1)}$ to (the pullback of)
the dual of the background six-form field strength $G_{(6)}$.

We assume a static (i.e. $y$-invariant)
$SO(4)$ symmetric configuration of the form $r(\theta)$,
$x(\theta)$ and $A_{y}(\theta)$ (with $\theta$ the polar angle of
the ${\bf S}^{4}$).
Using the explicit background\req{d4metric}
one can rewrite the Lagrangian (with a sign switch)
in the form
\be \label{u4}
U =  T_4 \Omega_{3}R^3\int d\theta
\sqrt{r^2+f(r)^{-1}r^{\prime 2} +(r/R)^{3}x^{\prime 2}}\,
\sqrt{D^2+\sin^6\theta}~,
\ee
where the displacement $D$ now satisfies the equation
$$
\partial_{\theta}D =-3\sin^3\theta,
$$
and is consequently given by
\be \label{d4}
D(\theta)=3\cos\theta-\cos^3\theta-2+4\nu~.
\ee
The constant of integration has been written again in terms of a
parameter $0\le\nu\le 1$, which controls the number of Born-Infeld
strings emerging from the D4-brane at each pole of the ${\bf S}^{4}$. 
For a
given $\nu$, it is easy to verify from\req{u4} that
the spikes at $\theta=\pi$ and $\theta=0$ will have the same 
asymptotic
`tension' as $(1-\nu)N$ and $\nu N$ fundamental strings, 
respectively.
It is clear then that $\nu$ is quantized in units of $1/N$ in the 
full theory.

{}From\req{u4} it is seen that if $\{r(\theta),x(\theta)\}$ is a 
solution
for horizon radius $r_{h}$, then
$\{\alpha r(\theta), \alpha^{-1/2}x(\theta)\}$
is a solution for horizon radius $\alpha r_{h}$. Notice that this
differs in functional form from what was seen in
Section \ref{eqn3dsec}. The scaling $x\propto
R^{3/2}/r^{1/2}$ is precisely
the holographic scaling expected from the UV/IR relation in the
near-horizon D4-brane background \cite{pp}.

After repeating essentially the same routine as in the threebrane 
case,
switching to the parametric representation $r=r(s)$, $x=x(s)$,
$\theta=\theta(s)$ one finds the constraint Hamiltonian
\be \label{ham4}
2\tilde{H} =  \left(1-\frac{r_h^3}{r^3}\right) p_r^2
  + \frac{p_{\theta}^2 }{ r^2} +
 \frac{p_x^2}{r^3}-
 \left( D^2(\theta)+\sin^6\theta \right)=0~.
\ee
The equations of motion have the same basic structure as those
obtained in the threebrane case:

\hfill
\parbox{1.7in}
{\begin{eqnarray*}
\dot r &=&\left(1-\frac{r_h^3}{r^3}\right) p_r~,\\
\dot\theta &=&\frac{p_\theta}{r^2}~, \\
\dot x &=& \frac{R^{3}}{r^{3}}p_{x},
\end{eqnarray*}}
\parbox{3.1in}
{\begin{eqnarray*}
\dot p_r &=&\frac{3}{2r^4}(p_x^2 R^3-p_r^2 r_h^3)
     + \frac{p_\theta^2}{r^3}~, \\
\dot p_\theta &=& -6\sin^3\theta \left(2\nu-1+\cos\theta
\right)~,\\
\dot{p}_{x}&=& 0~.
\end{eqnarray*}}
\hfill
\parbox{0.5in}
{\bea \label{eom4}
\eea}

Again, for large $r$ the Hamiltonian\req{ham4} reduces to that of a
particle moving in two dimensions, this time in a potential
$V(\theta)=-\left[D(\theta)^2+\sin^6\theta\right]$.
This potential has the same features as the one discussed in Section
3.1 and depicted in Fig.~1: two
minima at $\theta=0$ and $\theta=\pi$, and
a maximum at $\theta=\theta_{c}$ such that
$\cos\theta_c=1-2\nu$.

\subsection{Extremal Case: The BPS Baryon}

We begin the analysis of the solutions by specializing to
the extremal ($r_{h}=0$) background, with the aim of obtaining the
supersymmetric
fourbrane embeddings analogous to the fivebrane BPS configurations
found in~\cite{cgs}.
Following that paper, we set
$x^{\prime}=0$ in\req{u4}, to describe a baryon of zero-size.
The Euler-Lagrange equation for $r(\theta)$
is then
\be \label{el4}
{d\over d\theta}\left({r^\prime\over\sqrt{r^2+{r^\prime}^2}}
     \sqrt{D^2+\sin^6{\theta}} \right) =
 {r\over\sqrt{r^2+{r^\prime}^2}} \sqrt{D^2+\sin^6{\theta}}~.
\ee
The system under consideration here is in such close correspondence 
to
the one studied in~\cite{cgs}, that it is easy to
guess the BPS condition analogous to the one for a fivebrane embedded
in $AdS_{5}\times {\bf S}^{5}$ \cite{imamura,cgs,gomis}:
\be
\label{bps}
{r^\prime\over{r}} = {\sin^4\theta +D(\theta)\cos\theta
\over
     {\sin^3\theta\cos\theta -D(\theta)\sin\theta} }~,
\ee
where $D(\theta)$ is the function specified in\req{d4}. It is easy to
verify that\req{bps} is a first integral of\req{el4}. It is thus
almost certainly the condition for the fourbrane embedding to be
supersymmetric, although we have not checked this explicitly.

Just like its fivebrane cousin~\cite{cgs}, equation\req{bps}
can be solved analytically to obtain a
one-parameter family of BPS configurations,
\be
\label{bpssol}
r(\theta)=\frac{A}{\sin\theta}
          \left[\frac{\eta(\theta)}{2(1-\nu)}\right]^{1/2},
\qquad \eta(\theta)=1-2\nu-\cos\theta,
\ee
where the scale factor $A$ is arbitrary, and $\nu$ is the integration
constant in (\ref{d4}).
The solutions\req{bpssol} have the exact same features as their 
fivebrane
counterparts (see Eq. (9) and Fig.~1 in~\cite{cgs}):
they describe `tubes' of radius $A$
($r\sim A/(\pi-\theta)$ when $\theta\to\pi$),
with an asymptotic `tension' and total energy equal to those of
$(1-\nu)N$ strings.
The $\nu=0$ embedding has the shape of a `test tube'
with
$r(0)=r_{0}=A/2$. This solution captures all $N$ units of fourform
flux; it corresponds to a baryon in the $d=4+1$ SYM theory.
Notice that the surfaces\req{bpssol} are defined in the range where
$\eta(\theta)\ge 0$, namely
$\theta_{c}\le\theta\le\pi$,
where the critical angle $\theta_{c}$ is defined by
\be
\label{nu4}
\cos\theta_{c}=1-2\nu.
\ee
For $\nu>0$, the solutions intersect $r=0$ at $\theta=\theta_{c}>0$, 
and
capture only a fraction of the total flux. They represent gauge
theory objects with fewer than $N$ quarks (see the discussion
in~\cite{cgs}).

\subsection{Nonextremal Case: The Point Baryon}

Having understood the character of the BPS configurations in the
extremal background, we now
proceed to examine the solutions for finite horizon radius.
The fourbrane embeddings corresponding to baryons of the
four-dimensional gauge
theory with all of their component quarks at the same gauge
theory position ($x=0$)
and ${\bf S}^{5}$ angle ($\theta=\pi$), have $p_{x}=0$ (i.e.
$x^{\prime}=0$) and $\nu=0$ (see Section 3.4 for the case
$x^{\prime}\neq 0$, $\nu>0$).  As in Section 2.2, to obtain a smooth
solution we enforce the boundary
conditions $\partial_{\theta}r=0$ and $r=r_{0}$ at $\theta=0$.

Numerical integration with these boundary conditions yields results
completely analogous to those found for the three-dimensional case in
Section 2.2. There exists a
one-parameter family of solutions with parameter
$\zeta=r_{0}/r_{h}\ge 1$ (see Fig.~\ref{D3tubes}).
When
$\zeta<\zeta_{crit}\approx 18$ the solution tilts
towards and eventually crosses the symmetry axis,
thus reaching $\theta=\pi$ at a finite value of $r$. As $\zeta\to 1$
(i.e. $r_{0}\to r_{h}$),
the brane becomes more and more spherical (notice that
$r(\theta)=r_{h}$ is a solution to the equations of motion).
For $\zeta>\zeta_{crit}$, on the other hand,
the solution tilts away from the symmetry
axis, reaching $\theta=\pi$ only at $r=\infty$.

The cusp in the $\zeta<\zeta_{crit}$
solutions is again to be understood as the point of insertion of $N$
fundamental
strings (or equivalently, a collapsed fourbrane),
running along the ray $\theta=\pi$.
A variational calculation like the one performed in
Section~\ref{point3dsec}
shows that the stable configuration is
the one with $\zeta\to\infty$. So again the embeddings (similar to
those portrayed in
Fig.~\ref{D3tubes}) describe the process through which the $N$
fundamental strings deform the initially spherical fourbrane, pulling
it out to infinity. The final configuration has the shape of a 
`tube',
just like the BPS embeddings found in
the previous subsection. In the supersymmetric case
$r_{0}$ is a modulus of the configuration; but in the $r_{h}>0$ case
a potential drives the stable solution away to infinity.

\subsection{The Split Baryon: Color Dependence of the String Tension}
\label{split4dsec}

In this section we study solutions with
$p_{x}\neq 0$ (i.e. $x^{\prime}\neq
0$) and $\nu>0$, which correspond to a
baryon split into two separate groups of $\nu N$ and $(1-\nu)N$
quarks, a distance $L$ apart, and connected by a color flux tube.
The story here is in perfect parallel to the three-dimensional case
discussed in Section \ref{split3dsec}, although the quantitative
results are slightly different.
The two groups of
quarks are represented by Born-Infeld strings bundles
protruding from opposite poles of the ${\bf S}^{4}$,
$\theta\rightarrow 0$ and $\theta \rightarrow
\pi$.
The flux tube connecting the two bundles of quarks is dual to a
segment of the fourbrane that runs close and parallel to the horizon,
at the critical angle $\theta=\theta_{c}$ given by equation\req{nu4}.
This is the maximum of the
potential introduced at the end of Section \ref{eqn4dsec}.
The critical angle as a function of $\nu$ has the property that
$\theta_{c}(0) =0$, and $\theta_{c}(\nu)=\pi-\theta_{c}(1-\nu)$.

Numerical integration with initial conditions analogous 
to\req{inicon}
yields fourbrane embeddings of the same type as those found in the
three-dimensional case (see Figs.~\ref{split3d} and \ref{splitpol}).
For large quark separation $L$
the energy of the configuration is again
proportional to $L$, indicating confinement.
Due to the fact that the relation between $\nu$ and $\theta_{c}$ is
now simpler than that in Section \ref{split3dsec},
the tension of the flux tube can now be given very explicitly as a
function of $\nu$. It is
\be \label{tension4}
\sigma_{4}(\nu)=\frac{8\pi}{27}N(\g{4}^{2}N)T^{2}
      \nu(1-\nu)~.
\ee
The different powers of $T$, $N$, and the 't~Hooft
coupling $\lambda_{4}=\g{4}^{2}N$ match those found
in~\cite{groguri,baryonsugra,alisha}, where the baryon
is studied within a somewhat different scheme.

\begin{figure}[htb]
\medskip
\centerline{\epsfxsize=8cm
\epsfbox{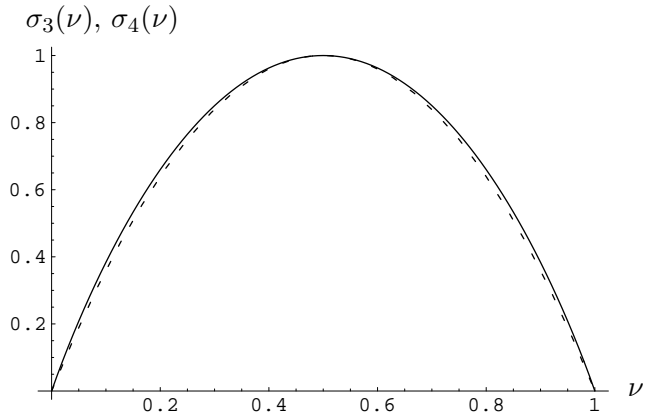}}
\begin{picture}(0,0)
\put(40,57){\small $\sigma_3(\nu)$, $\sigma_4(\nu)$}
\put(120,8){\small $\nu$}
\end{picture}
\caption{\small The tension of the flux tubes of the three-  and
 four-dimensional split baryon (solid and dashed curve, 
respectively),
 as a function of $\nu$, the fraction of separated quarks.
 See text for discussion.}
\label{D4tension}
\end{figure}

The simple dependence of the flux tube tension
on $\nu$ is perhaps what one would have naively guessed: the `product
of the charges'. It is amusing to see that this differs only slightly
from the three-dimensional result. For ease of comparison the
$\nu$-dependence of the tension (normalized to unity at the maximum)
is plotted for both cases in Fig.~\ref{D4tension}.

All the qualitative features of the tension as a function of $\nu$ 
are
the same as in the three-dimensional gauge theory.  In particular, 
for
$\nu=1/N\ll 1$ (i.e., when a single quark is pulled out), 
$\sigma_{4}$
is independent of $N$ (for given 't Hooft coupling) and exactly
matches the quark-antiquark string tension obtained
in~\cite{bisy2}.\footnote{Notice that equation~(3.20) in \cite{bisy2}
is incorrect by a factor of $\pi^3$.}

Finally, we want to emphasize that again the excitation energies 
associated with the flux tube tension\req{tension4} in the 
supergravity 
regime are larger than the compactification scale, 
so the gauge theory is not strictly four-dimensional 
\cite{groguri,bisy2}.

\section{Conclusions}

We have seen how studying the detailed shape of a D5/4-brane, 
embedded in
the near-horizon geometry of a large number $N$ of non-extremal
D3/4-branes, can reveal much information about the structure and
energetics of baryons in the corresponding strongly-coupled $SU(N)$ 
gauge
theory. In particular, we were able to construct embeddings 
representing a
baryon split into two separate clumps of quarks, with a color flux 
tube
running between them. As expected for a confining theory, the energy 
of
such a configuration is proportional to the separation between the 
two
quark bundles. Reading off the tension of the flux tube, we 
discovered a
non-trivial and physically reasonable dependence on the color charges 
of
the individual clumps.

We also studied the simpler `point-baryon' embeddings, where all the
quarks sit on top of each other.  The use of the full Born-Infeld 
plus
Wess-Zumino action for the system makes it clear that the naive 
picture of
the baryon as $N$ fundamental strings terminating on a spherical
five/four-brane is incomplete: the strings pull on the brane and 
deform
it. Our solutions displayed several interesting features, such as the
appearance of a potential for the overall scale of the system (which 
was a
modulus in the extremal case), that ultimately causes the brane to 
expand
out to infinity. 
Many
of the features in these solutions call out for a deeper explanation. 
In
particular, they point to the need for a fuller understanding of the
relation between the angular ($S^{5}/S^{4}$) dependence of the 
D5/4-brane
embedding and the flavor structure and energetics of the dual 
baryonic
state.

An obvious extension of our work would be to study the baryon in
thermal gauge theories. For the example of 
finite temperature (3+1)-dimensional SYM, one considers again
a fivebrane embedded in the nonextremal D3-brane
geometry\req{d3metric},
but this time extending along the compact $\tau$ direction.
A study of this system leads to some puzzling features, 
which we hope to clarify and report on in the near future.

Recently, Type 0 string theory has been suggested as a means to
construct gravity duals for large $N$ non-supersymmetric
four-dimensional gauge theories with phenomenologically interesting
behavior \cite{klebanov, minahan}. In particular, it has been argued
that, at least at the classical level, the Type 0 equations of
motion have asymptotic solutions which express asymptotic freedom in
the UV and confinement in the IR \cite{minahan}.  It would be very
interesting to study the baryon in this setting.  At present, the 
full
background geometry is not known, so the analysis would have to be
restricted to fivebrane embeddings which lie only in the asymptotic 
UV
region.  We hope that a complete study will eventually become
possible.

In our analysis the Born-Infeld string approach has again proven very
fruitful. We expect that many more lessons will be extracted with 
this
valuable tool in the future.

\section*{Acknowledgements}

This work was supported in part by US Department of Energy
grant DE-FG02-91ER40671 and by National Science Foundation grant
PHY98-02484. AG is also supported by the National Science and
Technology Council of Mexico (CONACYT).  AG would like to thank
I.~Klebanov and H.~Garc\'{\i}a-Compe\'an for
useful conversations. KGS is grateful to
G.~Savvidy for numerous helpful discussions.

\end{document}